\documentclass[aps,superscriptaddress,reprint,floatfix,prd]{revtex4-1}

\usepackage[plainpages=false, colorlinks=true, anchorcolor=blue, linkcolor=blue, citecolor=blue, bookmarks=false]{hyperref}
\usepackage{amsmath, amsthm, amssymb}
\usepackage{graphicx}

\usepackage{soul}
\usepackage{color}
\usepackage{dcolumn}
\usepackage{bm}
\usepackage[mathscr]{euscript}
\usepackage{mathrsfs}
\usepackage{amsbsy}

\begin{document}

\title{Axion driven cosmic magneto-genesis during the QCD crossover}

\author{F. Miniati}
\affiliation{Department of Physics, University of Oxford, Parks Road, Oxford OX1 3PU, UK}

\author{G. Gregori}
\email[Corresponding author: ]{gianluca.gregori@physics.ox.ac.uk}
\affiliation{Department of Physics, University of Oxford, Parks Road, Oxford OX1 3PU, UK}

\author{B. Reville}
\affiliation{School of Mathematics and Physics, Queen's University Belfast, Belfast BT7 1NN, UK}

\author{S. Sarkar}
\affiliation{Department of Physics, University of Oxford, Parks Road, Oxford OX1 3PU, UK}
\affiliation{Niels Bohr Institute, Blegdamsvej 17, 2100 Copenhagen, Denmark}

\begin{abstract}

We propose a mechanism for the generation of a magnetic field in the early universe during the QCD crossover assuming that dark matter is made of axions. Thermoelectric fields arise at pressure gradients in the primordial plasma due to the difference in charge, energy density and equation of state between the quark and lepton components. The axion field is coupled to the EM field, so when its spatial gradient is misaligned with the thermoelectric field, an electric current is driven. Due to the finite resistivity of the plasma an electric field appears that is generally rotational. For a QCD axion mass consistent with observational constraints and a conventional efficiency for turbulent dynamo amplification --- driven by the same pressure gradients responsible for the thermoelectric fields ---
a magnetic field is generated on subhorizon scales. After significant Alfv\'enic unwinding it reaches a present day strength of $B \sim 10^{-13 }$ G on a characteristic scale $L_B \sim $ 20 pc. The resulting combination of $BL_B^{1/2}$ is significantly stronger than in any astrophysical scenario, providing a clear test for the cosmological origin of the field through $\gamma$-ray observations of distant blazars. The amplitude 
of the pressure gradients may 
be inferred from the detection of concomitant
gravitational waves, while several experiments are underway to confirm or rule out the existence of axions.
%
\end{abstract}

\maketitle

Magnetic fields are a ubiquitous feature of astrophysical plasmas and may plausibly have originated from primordial seed fields~\cite{1983flma....3.....Z}. In fact the `turbulent dynamo' can efficiently amplify tiny, i.e.  dynamically negligible, magnetic seeds at an exponential rate~\cite{2005PhR...417....1B,Schekochihin_Turbulence_2006}, a result supported by laboratory experiments~\cite{Meinecke_Turbulent_2014,Meinecke_Developed_2015,Gregori_The_2015}. 
Initial seeds are nevertheless nontrivial to generate in a perfectly conducting astrophysical plasma. Viable mechanisms proposed earlier include Compton drag~\cite{1970MNRAS.147..279H,2006Sci...311..827I} or the `Biermann battery', both at cosmic shocks and ionisation fronts~\cite{1994MNRAS.271L..15S,1997ApJ...480..481K,Gregori_Generation_2012},
and speculative processes in the early universe~\cite{2007NewAR..51..275D}.
In addition for most astronomical systems --- from compact objects like stars to the interstellar medium of galaxies --- the time scale of the turbulent dynamo (in the regime when the magnetic field is dynamically important) is much shorter than the age of the universe and, in agreement with astronomical observations, the growth of magnetic energy saturates as it reaches approximate equipartition with the turbulent kinetic energy. This picture for the turbulent dynamo origin of magnetic field appears to continue to hold even in galaxy clusters, where the scale of the turbulence is much greater and consequently the dynamo remains unsaturated~\cite{Miniati:2015jj,2016ApJ...817..127B}.

However, recent analyses of the $\gamma$-ray spectra of distant blazars suggest the presence of a   magnetic field exceeding $10^{-18}-10^{-15}$~G in cosmic voids~\cite{2010Sci...328...73N,2010MNRAS.406L..70T,2011MNRAS.414.3566T} where the turbulent dynamo is \emph{unlikely} to operate. The only viable astrophysical mechanism for the generation of such fields is based on return 
currents induced by cosmic-rays escaping from young galaxies and streaming in the high resistivity, non-uniform intergalactic plasma, just prior to reionisation~\cite{2011ApJ...729...73M,2012ASPC..459..125M,Durrive_Intergalactic_2015}. This mechanism cannot however generate a field stronger than $\sim 10^{-17}-10^{-16}$~G, while the rather conservative lower limit of 10$^{-18}$ G from current blazar observations~\cite{2011ApJ...733L..21D} is likely to improve significantly when the Cherenkov Telescope Array begins operations~\cite{2016ApJ...827..147M}. It is of interest, therefore, to explore alternative scenarios of cosmic magnetogenesis.

The early universe provides perhaps the only alternative scenario for the generation of magnetic fields in cosmic voids. The quantum-chromodynamic (QCD) transition which occurs at $t_\mathrm{QCD}\sim 10^{-5}$ s, $T_\mathrm{QCD}\sim 150$ MeV, when free quarks and gluons are confined into hadrons, is particularly interesting as thermoelectric fields arise at pressure gradients as a result of the different charge density, energy density and equation of state of the quark and lepton  components~\cite{Quashnock_Magnetic_1989}. It was previously thought that the confinement process involves a first-order phase transition resulting in hadronic bubbles that nucleate and collide producing shock waves throughout the primordial plasma~\cite{Banerjee_Evolution_2004,Pen_Shocks_2016}. Under these conditions the thermoelectric field would have a rotational component and, analogous to the case of the `Biermann battery', generate a magnetic field~\cite{Quashnock_Magnetic_1989,Brandenburg_Large_1996}.  However, it has subsequently been realised via realistic lattice simulations that the confinement of hadrons does \emph{not} involve a phase transition but is just a smooth crossover without formation of shocks~\cite{Aoki_The_2006,Rischke_The_2004}. The fluid remains barotropic and the thermoelectric field irrotational.  However the latter may still generate a magnetic field~\cite{1992PhRvD..46.5346G} by interacting with a pseudo-scalar axion field~\cite{doi:10.1146/annurev.nucl.012809.104433}, which couples to the electro-magnetic (EM) field via the Primakoff mechanism. This is the mechanism we consider in this paper, assuming that dark matter in fact consists of axions. 

The axion ($a$) is the pseudo Nambu-Goldstone boson of the broken $U(1)$ Peccei-Quinn symmetry \cite{Peccei_CP_1977} which was introduced to explain the absence of $CP$ violation in strong  interactions~\cite{PhysRevLett.40.223,PhysRevLett.40.279}. It acquires a mass $m_a \simeq 6~\mu\mathrm{eV} (10^{12} \, \mathrm{GeV}/f_a)$ at  temperatures below the QCD scale through mixing with the $\pi^0$ and $\eta$ mesons. Here $f_a$ is the axion decay constant, related to the vacuum expectation value $v_a$ that breaks the Peccei-Quinn symmetry as $f_a = v_a/N$, where $N$ characterises the colour anomaly of $U(1)_\mathrm{PQ}$ and equals 6 for the original Weinberg-Wilczek-Peccei-Quinn axion model. If cosmological inflation occurs \emph{after} the Peccei-Quinn symmetry breaking, the axion field is homogenised over the observable universe~\footnote{Note that even when the reheating temperature after inflation $T_\mathrm{R}$ exceeds $v_a$ thus restoring the Peccei-Quinn symmetry, topological defects such as domain walls and strings form in the axion field and these too can give rise to magnetic fields as they decay~\cite{Field_Cosmological_2000,Forbes_Primordial_2000}. However, non-zero helicity and turbulence are then needed in the early universe for the magnetic field to grow,  and this is \emph{not} consistent with CMB data.}. When the temperature drops to the QCD scale the axion field acquires a mass and starts oscillating on a scale $\lambda_{a}=2\pi/m_{a}$. The energy density of these oscillations is of order the critical density of the universe for $f_a \sim 10^{12} \,\mathrm{GeV}$, hence the ``invisible axion" is a well-motivated candidate for the dark matter~\cite{PRESKILL1983127,willy,ABBOTT1983133,2008LNP...741...19S,2008LNP...741...19S}.
The axion mass effectively sets in at $T \sim 1$ GeV ($t \sim 2\times 10^{-7}$ s) due to non-perturbative QCD instantons which turn on sharply as the temperature drops, hence from then onwards the axion field oscillates around a constant value~\cite{2008LNP...741...19S}. Accordingly, the axion field evolves as  $a(R)=(2\rho_\mathrm{DM}^0)^{1/2}/R^{3/2} m_a$~\cite{2013arXiv1309.7035C},
with $\rho_\mathrm{DM}^0=9.6\times 10^{-12}$ eV$^4$ the comoving dark matter density~\cite{1674-1137-40-10-100001} (hereafter we use natural units $\hbar=c=k_B=1$), and
$R$ the scale factor of the universe normalized to unity at the present time~\footnote{This scaling with $R$ follows because the equation of state of the oscillating axions is that of an ideal gas of non-relativistic particles~\cite{2008LNP...741...19S}}.

The axion coupling to the EM field is commonly described through a Lagrangian term $\mathcal{L}_{\rm int} = -g_{a \gamma} {\bf E} \cdot {\bf B}\,a$, where
$g_{a \gamma} \equiv \alpha\xi/\pi f_a=\xi \times 10^{-22} (m_a/{\rm meV})$ eV$^{-1}$ is the axion-photon coupling, $\alpha$ is the fine structure constant, $\xi \sim 2$ depending
on the specific axion model considered \cite{Duffy_Axions_2009},
and ${\bf E},~{\bf B}$ are the electric and magnetic field, respectively.
With the addition of the above Lagrangian term, Maxwell's equations in comoving coordinates read~\cite{Harari_Effects_1992,1992PhRvD..46.5346G}:
\begin{eqnarray}
\nabla \cdot {\bf E} &=& \rho +\frac{g_{a \gamma}}{R^{3/2}} {\bf B} \cdot 
\nabla a,\\
\nabla \cdot {\bf B} &=& 0, \\
\nabla \times {\bf B} &=& {\bf J} + R\frac{\partial {\bf E}}{\partial t} + \\ \nonumber & &
\frac{g_{a \gamma}}{R^{3/2}} 
\left[{\bf E} \times \nabla a - R{\bf B}\left(\frac{\partial a}{\partial t}-\frac{3}{2}\frac{\dot R}{R}a\right)\right], \\
\nabla \times {\bf E} 
&=&-R\frac{\partial {\bf B}}{\partial t},
\end{eqnarray}
where $\rho$ and $\bf J$ are the charge and current density, respectively, 
and all variables are in comoving units.
In particular the $\bf E$ and $\bf B$ fields are 
subtracted off the respective components arising 
from Hubble expansion.

As mentioned already, a thermoelectric field arises at pressure gradients in the primordial plasma due to a slight asymmetry in the
charge, energy density and equation of state of the quark and leptonic components. Thus the strength of the field is only a fraction $\epsilon \sim 0.1$ of the usual baroclinic term~
\cite{Quashnock_Magnetic_1989}.
Hence, in the absence of a magnetic field,
Ohm's law in comoving units reads 
\begin{equation}
{\bf E} = \eta_p {\bf J}-\epsilon \frac{\nabla P}{e n}, 
\label{ohm}
\end{equation}
where $\eta_p = \pi e^2/RT$ \cite{Durrer:2013ec} is the comoving plasma resistivity (for $T>1$ MeV), $P = 7 \pi^2 g^* R^4T^4/720$ the 
comoving pressure with $g^*$ the number of relativistic degrees of freedom, 
$n=3\zeta(3) g^* R^3T^3/4\pi^2$ the comoving density with $\zeta$ the Riemann zeta function (and $\zeta(3) \approx 1.2$). We notice that while $g^*$ can vary up to an order of magnitude near the QCD crossover, the electric field remains \emph{insensitive} to its
value. Taking again the magnetic field to be initially zero, Ampere's equation yields
\begin{equation}
\label{Jeq}
{\bf J} \approx -\frac{g_{a \gamma}}{R^{3/2}}
\left({\bf E} \times \nabla a \right).
\end{equation}
Substituting the above into Ohm's law (\ref{ohm}) we find an electric field
\begin{equation}\label{Eeq}
{\bf E}=-
\frac{{\pmb{\mathcal{A}}} ({\pmb{\mathcal{A}}} \cdot {\pmb{\mathcal{H}}}) + 
{\pmb{\mathcal{A}}} \times {\pmb{\mathcal{H}}} + 
{\pmb{\mathcal{H}}}}{1+{\mathcal{A}}^2} ,
\end{equation}
%
where,
\begin{equation}
\label{AHeq}
{\pmb{\mathcal{A}}} = \eta_p 
\frac{g_{a \gamma}}{R^{3/2}} \nabla a,\quad
{\pmb{\mathcal{H}}} = \epsilon \frac{\nabla P}{e n}.
\end{equation}
Using $\lambda_a$ as the characteristic scale length of the axion field,
we can determine $\pmb{\mathcal{A}}$ (with $\xi=2$):
\begin{equation}
\mathcal{A} \approx 1.4\times 10^{-19}
\left(\frac{T_\mathrm{QCD}}{\rm 150~MeV}\right)^{-1}
\left(\frac{m_a}{\rm meV}\right)
\left(\frac{R}{R_\mathrm{QCD}}\right)^{-\frac{3}{2}}.
\end{equation}
Thus, the electric field is dominated by the thermoelectric term, $\pmb{\mathcal{H}}$.
When $\pmb{\mathcal{A}}=0$, the curl of the electric field describes the generation of magnetic field via the usual baroclinic mechanism. This, however, vanishes in the present context 
as the fluid remains barotropic during the QCD crossover.
The thermoelectric field plays nevertheless an important role. Inspection of Eqs.(\ref{Jeq}) and (\ref{Eeq}) shows that, provided the axion field gradient and the thermoelectric field are not exactly aligned, an electric current is driven in the primordial plasma through their interaction. Owing to the finite resistivity of the plasma the current has an associated electric field, which gives rise to the new terms in Eq.(\ref{Eeq}). Unlike the simple thermoelectric term this resistive field has in general a \emph{rotational} component. In view of the smallness of $\pmb{\mathcal{A}}$ the rotational electric field is simply the cross product $\pmb{\mathcal{A}}\times \pmb{\mathcal{H}}$.

This process generates a magnetic seed of strength  $\mathcal{A}\mathcal{H}\,t_{\rm QCD}\sim 0.1 \,\mu$G. However, the same pressure gradients giving rise to the thermoelectric fields will generally drive large scale plasma motion initiating a turbulent cascade. This can lead to significant amplification of the initial seed by turbulent dynamo action.
The amplitude of the pressure fluctuations is 
potentially observable via the detection of the associated emission of gravitational waves~\cite{MouraoRoque:2013fb}. Nevertheless, detailed studies of the QCD crossover on the lattice are consistent with expectations from the phenomenological hadron resonance gas model in which fluctuations in the thermodynamic properties of order $\sim 1/\sqrt{g_*}$ arise on all scales up to the horizon \cite{qcd1,qcd2,qcd3}. This implies fluctuations, $\delta P/P$, of similar order in the pressure of the
hadronic plasma which in turn induce velocity fluctuations of strength $\delta u/c_{\rm s} \sim 1/\sqrt{g_*}$, where $c_{\rm s}=1/\sqrt{3}$ is the sound speed. 

We expect such velocity fluctuations to stir up a turbulent cascade if the eddy-turnover rate is faster than the Hubble expansion, i.e. up to a scale $L_u\approx \delta u L_H \approx 0.1\,(g^*/30)^{-1/2}\,L_H$, with $L_H=2\,t_{\rm QCD}$ the particle horizon at that time. Numerical simulations show that the power spectrum of the turbulent cascade for mildly relativistic flows, i.e. with Lorentz factor $\Gamma \simeq 0.4-1.7$ appropriate for the velocity fluctuations considered here, is well described by the classical theory of Kolmogorov~\cite{2009ApJ...692L..40Z,2013ApJ...766L..10R}. Since the induction equation remains unchanged in the relativistic regime, we expect the turbulent dynamo to operate analogously to the classical case ~\cite{2005PhR...417....1B,Schekochihin_Turbulence_2006}, as is supported by numerical studies that have started to address such questions~\cite{2009ApJ...692L..40Z,2014MNRAS.439.3490M}.
In particular the magnetic field is expected to grow exponentially and thus quickly reach equipartition with the turbulent kinetic energy at the Kolmogorov scale during the initial kinematic phase, and grow thereafter at a rate that is a fraction $\eta_{B}$ of the turbulent dissipation rate, $\varepsilon_{\rm turb}\simeq (1/3)^{3/2}\rho\,\delta u^3/L_u$~\cite{2012MNRAS.422.3495B,2016ApJ...817..127B}, where the radiation energy density is $\rho=\pi^2g^*T^4/30$. The magnetic field growth carries on over an e-folding time, $t_{\rm QCD}$, stalling afterwards when the large-scale velocity field is damped by cosmological expansion. The magnetic energy accumulated in the process is thus
\begin{equation}
E_B \approx \left(\frac{1}{3}\right)^{3/2} \eta_B\, \rho\frac{\delta u^3}{L_u}t_{\rm QCD}
\approx  0.1\,\eta_B\,\pi^2 \frac{T^4}{90},
\end{equation}
independent of $g^*$. Since $L_u/\delta u\approx 2\,t_{\rm QCD}$, this is simply a fraction $0.2\,\eta_B$ of the available kinetic energy. For an efficiency $\eta_B$ of order a few percent, characteristic of very high Reynolds number flows ~\cite{2012MNRAS.422.3495B,2016ApJ...817..127B}, the magnetic field in comoving units is
\begin{equation} \label{BQCDeq}
B_{\rm QCD} \approx 9 \times 10^{-8}\,\left(\frac{T_{\rm QCD}}{150~{\rm MeV}}\right)^2 \left(\frac{\eta_B}{0.05}\right)^{1/2} {\rm G},
\end{equation}
yielding an Alfv\'en speed
\begin{equation} \label{Alfveneq}
v_{\rm QCD} =\frac{B}{\sqrt[]{\rho}}=10^{-2}\left(\frac{\eta_B}{0.05}\right)^{1/2} 
\left(\frac{g^*}{30}\right)^{-1/2},
\end{equation}
and a characteristic Alfv\'en scale, where the magnetic energy balances the turbulent kinetic energy
(assuming again a Kolmogorov cascade):
\begin{equation} \label{LBQCDeq}
L_{\rm QCD} \approx \frac{\eta_B^{3/2}}{9}\, L_u
\approx 10^{-4}\left(\frac{\eta_B}{0.05}\right)^{3/2}
\left(\frac{g^*}{30}\right)^{-1/2}\, L_H.
\end{equation}
As the large scale velocity flow is damped by cosmological expansion and the cascade dissipates,
the dynamo action stops and the magnetic field begins to unwind. This roughly leads to a configuration in which the magnetic field is coherent on domains of order the Alfv\'en scale and uncorrelated on larger scales. As the field unwinds the magnetic tension at the boundaries of these domains will tend to rearrange the field at a rate determined by the Alfv\'en speed in such a way as to increase its correlation length, freeing magnetic energy in the process. If the magnetic field is frozen in the plasma, the correlation length will simultaneously increase due to cosmological expansion. The growth of the magnetic field correlation length can thus be described by the following equation~\cite{Dimopoulos:1997wx} conveniently cast in comoving form as:
\begin{equation} \label{LBeq}
\frac{\rm d}{{\rm d} t}\frac{L_B}{R}=\frac{v_{\rm A}}{R}.
\end{equation}
The release of magnetic energy associated with the realignment of randomly oriented domains of coherent field is governed by conservation of magnetic flux. This assumption is justified below where it is found that effects due to finite resistivity of the plasma may safely be neglected. For a magnetic field $B$ correlated up to a scale $L_B$, conservation of magnetic flux from randomly superposed magnetic bundles through a surface enclosed by a material line of size $L\gg L_B$, implies $BL^2\propto L/L_B$. Furthermore, if the plasma in which the field is frozen is subject to cosmological expansion, $L\propto R$, one has
\begin{equation} 
\label{BCOeq}
B\, R^{2}\propto\frac{R}{L_B},
\end{equation}
so that a comoving magnetic field will decay during cosmic expansion if its correlation length, $L_B$, grows faster than the scale factor~\cite{Dimopoulos:1997wx}. This description for the coupled evolution of $B$ and $L_B$ for a decaying non-helical turbulent magnetic field based on Eqs.(\ref{LBeq}) and (\ref{BCOeq}) appears to really capture the essential physics despite its simplicity. In fact, ignoring for a moment the effect of cosmic expansion (i.e. setting $R,\rho={\rm const}$), Eq. (\ref{BCOeq}) implies the constancy of the magnetic Lundquist number, $Lu_{\rm M} \propto v_{\rm A} L_B \propto BL_B$. Thus $v_{\rm A} \propto 1/L_B$ and substituting into Eq.(\ref{BCOeq}) and solving we find $L_B\propto t^{1/2}$. This result and the constancy of $Lu_{\rm M}$ are indeed fully consistent
with results from numerical simulations of decaying non-helical magnetic fields in incompressible turbulent flows, both in the classical~\cite{Brandenburg:2015dn} and relativistic regime~\cite{2014ApJ...794L..26Z}.

Apart from the effect of cosmic expansion, the evolution of $L_B$ depends on the Alfv\'en speed, $v_{\rm A}=B/\sqrt[]{\rho}$. During the radiation-dominated era the energy density $\rho\approx\rho_{\rm rad}$ decreases as $R^{-4}$. Thus taking into account Eq.(\ref{BCOeq}), the Alfv\'en speed evolves as
\begin{eqnarray} 
\label{VAeq1}
\frac{v_{\rm A}}{v_{\rm QCD}} =
 \frac{R}{R_{\rm QCD}} 
 \frac{L_{\rm QCD}}{L_{B}} .
\end{eqnarray}
Substituting this into Eq.~(\ref{LBeq}) and integrating (analytically) shows that indeed magnetic tension causes a growth of the correlation length, $L_B \propto R^{3/2}$, significantly faster than the Hubble expansion ($\propto R$). This solution remains valid until the photon mean free path for Thomson scattering, $\ell_{\rm T}=1/\sigma_{\rm T}n_e\propto R^3$, with $\sigma_{\rm T}$ the Thomson cross section and $n_e$ the number density of free electrons/positrons, becomes larger than $L_B$. 
For the parameters describing our problem this transition takes place around $R_{\rm T}\approx 4\times 10^{-7} (\eta/0.05)^{2/3}$. At this point radiation drag, $F_{\rm drag}=\rho v_{\rm T}/\ell_{\rm T}$, effectively inhibits the plasma motions induced by magnetic tension, $B^2/L_B$, and the unwinding of the field lines proceed at the much slower terminal speed~\cite{1983PhRvL..51.1488H} 
\begin{equation}
v_{\rm T}=v_{\rm A}^2\frac{\ell_{\rm T}}{L_B}.
\end{equation}
This situation persists beyond the epoch of radiation-matter energy density equality
until shortly after the (re)combination era, $R_{\rm rec} \approx 10^{-3}$, when neutral hydrogen forms and the fraction of free electrons drops dramatically to  $x_e\approx 2.3\times 10^{-4}$, causing an increase in the photon mean free path by a factor $x_e^{-1}$. The evolution equation for the correlation length (\ref{BCOeq}) with $v_{\rm A}$ properly replaced by $v_{\rm T}$ can again be integrated analytically, resulting in a correlation length at recombination
\begin{equation}
L_{\rm rec} \approx  \frac{8}{3} v_{\rm T} (t_{\rm rec}) t_{\rm rec}
\approx 10^{-3}\left(\frac{\eta_{B}}{0.05}\right) \left(\frac{g^*}{30}\right)^{-\frac{1}{2}}\, {\rm pc},
\end{equation}
where $t_{\rm rec}$ is the time at recombination. The magnetic field strength according to Eq.(\ref{BCOeq}) is then
\begin{equation} 
B_{\rm rec} \approx B_{\rm QCD} \frac{L_{\rm QCD}}{R_{\rm QCD}}  \frac{R_{\rm rec}}{L_{\rm rec}} \approx 10^{-12} \left(\frac{\eta_{B}}{0.05}\right)\,{\rm G}.
\end{equation}
After recombination the baryonic fluid is no longer coupled to the radiation field so that radiation drag is ineffective. In addition, the inertia of the plasma is determined by the baryonic matter alone: $\rho_{\rm b}\propto R^{-3}$. As a result the Alfv\'en speed is boosted by a factor $(\Omega_{\rm m}/\Omega_{\rm b})^{1/2}$, the square root of the ratio of total to baryonic matter density, and its growth with the cosmological expansion slows to being $\propto R^{1/2}$. Compared to Eq.(\ref{VAeq1}) the evolution of the Alfv\'en speed is therefore rescaled as:
\begin{equation} 
\label{VAeq2}
\frac{v_{\rm A}}{v_{\rm QCD}} =
\left(\frac{\Omega_{\rm m}}{\Omega_{\rm b}}\right)^{1/2}\left(\frac{R_{\rm eq}}{R_{\rm QCD}}\right)^{1/2} \left(\frac{R}{R_{\rm QCD}}\right)^{1/2} \frac{L_{\rm QCD}}{L_B}.
\end{equation}
Using this expression to integrate Eq.~(\ref{BCOeq}) we finally find the correlation length at present,
\begin{equation} 
\label{L0}
L_{0}  \approx 17\,v_{\rm A}\,(t_H)\,t_H
\approx  25 \left(\frac{\eta_{B}}{0.05}\right)\left(\frac{g^*}{30}\right)^{-1/2}\, {\rm pc},
\end{equation}
where $t_H$ is the Hubble time. The   magnetic field strength is:
\begin{equation}  
\label{B0}
B_{0} \approx  5\times 10^{-14} \left(\frac{\eta_{B}}{0.05}\right)\, {\rm G}. 
\end{equation}

Concerning our assumption of the conservation of magnetic flux, we note that the resistive scale corresponding to $\lambda_\eta\approx 3.4\times 10^{-5} (R/R_\mathrm{QCD})^{3/2}$ cm for $T\geq 1~$MeV~\cite{Dimopoulos:1997wx} (and taking Spitzer's value in the nonrelativistic regime at lower temperature) remains much smaller than the magnetic field correlation length, $L_B$, both in the early universe and up until reionisation, when the temperature of the intergalactic plasma is smallest (and consequently the resistive scale largest). Therefore the characteristic scale of the magnetic field is not affected by this process.

The magnetic field strength in Eq.(\ref{B0}) depends essentially on a single parameter, the assumed efficiency of dynamo action, $\eta_B$. This important parameter appears to converge to a value of 0.05 (corresponding to a mean value for time dependent flows) in the limit of high Reynolds number flows~\cite{2012MNRAS.422.3495B,2016ApJ...817..127B} and in future may also be determined experimentally \cite{tzeferacos2018laboratory}. In any case our result rests upon the assumption that the velocity perturbations generated during QCD crossover scale as $1/\sqrt{g^*}$, as indicated by lattice simulation studies of this process~\cite{qcd1,qcd2,qcd3}, and possibly testable in the future through the detection of the associated emission of gravitational wave (eLISA/New Gravitational Wave 
Observatory~\cite{2012JCAP...06..027B}, Big Bang Observatory, TOBA~\cite{Ishidoshiro:2011ky}).
The magnetic correlation length shows an additional mild dependence on (the inverse square root of) the number of relativistic degrees of freedom of the plasma, which can always be accurately  calculated numerically.
The magnetic field strength and its correlation length which we have found are rather insensitive to the exact value of the axion wavelength as long as $\lambda_a \ll L_{\rm QCD}$, since the eddy-turnover time $L_{\rm QCD}/\delta u\approx 10^{-3}t_{\rm QCD}$ is significantly shorter than the Hubble time at the QCD crossover. This can be expressed as a constraint on the QCD axion mass, $m_a \gg 2.5\times 10^{-3}$ meV, which can be experimentally tested~\cite{Rosenberg_Dark_2015} and
is consistent with the estimated range $\sim 0.05-1.50$~meV corresponding to  the assumption that axions contribute between 100\% and 1\% of the dark matter (in the post-inflation scenario)~\cite{Borsanyi_Calculation_2016}.

From the observational point of view, for any reasonable choice of the turbulent dynamo efficiency, 
$\eta_B \lesssim 0.05$, the magnetic  field strength is safely below the constraint imposed by Big-bang nucleosynthesis, as well as the more stringent limit of  $10^{-9}$ G set by the observed CMB anisotropy~\cite{1983flma....3.....Z,Durrer:2013ec,2000PhRvD..61d3001D}. It is also compatible with the upper bound of $\sim 3 \times 10^{-9}$ G set by the maximal magnetic pressure support compatible with the formation of small cosmic structures~\cite{Schleicher_Primordial_2011}.
Finally, the magnetic field strength and correlation length yield the following observable combination
\begin{equation} \label{BLeq}
\left(\frac{B_0}{\rm G}\right)\left(\frac{L_0}{\rm kpc}\right)^\frac{1}{2}
\approx 10^{-14}
\left(\frac{\eta}{0.05}\right)^\frac{3}{2}
\left(\frac{g^*}{30}\right)^{-\frac{1}{4}}.
\end{equation}
This quantity is of interest because it determines the degree to which the magnetic field in voids affects observable properties of the secondary cascade emission initiated by multi-TeV photons from distant blazars. These include  modification of the spectral energy distribution, broadening of the angular profile and the time-delay correlation in blazar radiation at different energies. In particular, the observed absence of a GeV bump expected in the spectra of a number of distant blazars due to the absorption and reprocessing of their multi-TeV emission is interpreted as evidence of a magnetic field in voids stronger than $8\times 10^{-16}$ G~\cite{Neronov_Evidence_2010, Taylor:2011bx, 2011ApJ...733L..21D}. The strength predicted by axion magnetogenesis is significantly higher than this, in contrast to astrophysical models~\cite{2011ApJ...729...73M,Durrive_Intergalactic_2015}. There are other observable distinctions, e.g. in contrast to astrophysical models the magnetic field generated in our model is strong enough to cause a broadening of the secondary cascade emission which, although presently unobserved~\cite{HESSCollaboration:2014ej,2014ApJ...787..155T,Chen:2015bq,Archambault:2017cn}, should be  detectable with CTA \cite{cta}.

{\bf Acknowledgements:}
This research was supported by the UK Engineering \& Physical Sciences Research Council (EP/M022331/1, EP/N014472/1, EP/N002644/1), the UK Science \& Technology Facilities Council (ST/P000770/1), and the Danish National Research Foundation (DNRF91). We thank Jorge Casalderrey Solana for a helpful discussion concerning fluctuations during the QCD crossover, and Konstantin Beyer for checking our calculations.

\end{document}